# Geoscience and the Search for Life Beyond the Solar System

Rory Barnes
University of Washington
rory@astro.washington.edu
206-543-8979

Anat Shahar (Carnegie Institute for Science)
Cayman Unterborn (Arizona State University)
Hilairy Hartnett (Arizona State University)
Ariel Anbar (Arizona State University)
Brad Foley (Penn State University)
Peter Driscoll (Carnegie Institute for Science)
S.-H. Dan Shim (Arizona State University)
Thomas Quinn (University of Washington)
Kayla Iacovino (Arizona State University)
Stephen Kane (University of California-Riverside)
Steven Desch (Arizona State University)
Norman Sleep (Stanford University)
David Catling (University of Washington)

# I. Introduction

How can scientists conclude with high confidence that an exoplanet hosts life? As telescopes come on line over the next 20 years that can directly observe photons from terrestrial exoplanets, this question will dictate the activities of many scientists across many fields. The expected data will be sparse and with low signal-to-noise, which will make disentangling biosignatures from abiotic features challenging. Our Earth is not just unique in that it hosts life, it is also the only terrestrial planet with direct observations of its interior through seismic waves, and compositional evolution through field and laboratory measurements. This extensive research reveals a planet born from collisions between worlds (Canup & Asphaug, 2001), followed by a complicated biogeochemical evolution (Lyons et al., 2014). Exoplanet interiors, on the other hand, can only be constrained by the following observations: 1) photometric and spectroscopic analysis of the planet's atmosphere, 2) spectroscopic and photometric analysis of the host star, and 3) companion planet properties. From these (future) data, astrobiologists must generate plausible compositional and evolutionary models that constrain a potentially habitable exoplanet's internal properties and history, provide environmental context, and rule out geochemical explanations for any putative biosignatures. The goal of this white paper is to frame the role of geophysical and geochemical processes relevant to the search for life beyond the Solar System and to identify critical, but understudied, areas of future research.

The emerging field of "exogeoscience" is the study of how galactic, stellar system, atmospheric, and internal processes of terrestrial exoplanets affect the properties, evolution, and observable features of their surfaces and interiors. These phenomena and their couplings are central to the concept of planetary habitability, an environmental state that permits the origination and sustainment of life, because all plausible theories require a solid surface under a liquid water layer. As biospheres sit atop a tectonically active solid planet that can generate a magnetic field above the atmosphere, solid body processes are fundamental to both theoretical models (Foley & Driscoll 2016) and retrieval algorithms (Meadows et al. 2018). Yet the challenge of measuring internal properties remotely, e.g. with photometric and/or spectroscopic data from future space- and ground-based facilities, is profound: Exoplanets are too distant for robotic exploration, and solid surfaces are opaque. Without significant investment of resources in theoretical and laboratory research to understand the full range of interior processes on exoplanets, interpreting spectral features as biosignatures will be purely speculative. Below we describe the current state of exogeoscience, and then suggest research initiatives that could dramatically improve the chances of unambiguously identifying active biology on an exoplanet.

# II. Current Observations

Earth observations and analyses span field investigations, lab experiments, and theoretical work. A proper summary of Earth science is too long to review in this format, but Earth, suffice to say, is an extremely complex system with an equally complex history that is still being pieced together by geophysicists, geochemists, planetary scientists, atmospheric scientists and astrophysicists. This effort must advance beyond explaining Earth data in order to develop general principles that can be used to explain – and ultimately predict – observations from exoplanets that differ from Earth in mass, size, and chemical makeup.

Terrestrial exoplanets have been found orbiting a wide range of stars with a wide range of orbits. For stars about the size of the Sun and larger, asteroseismology can provide mass, radius and age measurements accurate to about 10% that can, in turn, provide important constraints on the bulk planet properties. Stellar spectra can provide relative abundances of elements, but isotopic abundances are only available for a small number of stars that are very similar to our Sun. Most stars form in small "embedded clusters," but about 10% form in large clusters with nearby supernovae (Lada & Lada, 2003; Fatuzzo & Adams, 2015). The recent kilonova explosion GW081717 has shown that neutron star-neutron star mergers form significant amounts of heavy elements far from planet forming regions (Abbott et al., 2017). A recent survey of more than 1000 FGK stars in the Galaxy demonstrated a factor of two variation in the major element composition (e.g., Mg/Si and Fe/Si) of these stars (Adibekyan et al., 2015; Delgado Mena et al., 2010). These ratios determine the size of the planet's core, the mineralogy, melting temperature, viscosity, conductivity of the mantle, storage capacity for volatiles, heat producing elements, etc. Minor elements change heating rates and crustal compositions (Unterborn et al., 2016; 2017a). Terrestrial exoplanets could have a wide range of compositions, radiogenic abundances, initial temperatures, tidal heating, and orbit in systems with orbital architectures very different from our own. An integrated, exogeoscience approach is thus needed to both assess the habitability of these planets, as well as to identify key diagnostics that differentiate inhabited, habitable, and sterile worlds.

## III. The Exogeoscience Framework

The initial conditions of a terrestrial planet are set by its local environment at the time of formation, whereas the observed state of a planet is determined by its formation *and* subsequent evolution. Galactic chemical evolution increases the abundances of heavier elements, which affect molecular cloud compositions, the protoplanetary disks, and ultimately the compositions of the planets themselves. Collisions, abundances of radioactive isotopes, and tidal heating set the initial thermal state. The evolution of the planet's interior and surface are controlled by the dynamical evolution of the interior: the movement of heat and mass through the surface, the atmosphere, and into space. Figure 1 shows a schematic of these connections. Considerable research has addressed multiple aspects of these phenomena and points toward a much greater diversity of terrestrial exoplanets than is present in our Solar System (e.g. Léger et al. 2004; Bond et al., 2010; Frank et al. 2015; Luger & Barnes 2015).

Galactic models find that heavy element abundance increases with time and proximity to the galactic center (see, e.g., Schönrich & Binney, 2009). As galactic dynamics permits stars to travel radially through the galaxy by over 10,000 light years (Sellwood & Binney, 2002), exoplanets in the stellar neighborhood may have formed in very different environments. The galactic center and supernova remnants also generate cosmic rays that may alter the atmosphere of a terrestrial exoplanet if they pass near them.

As stars form, a disk of material naturally develops around them that can produce planets. The abundances of solids depend on orbital distance, composition, temperature and pressure, which can be modified by stellar, star cluster, and protoplanet effects. Planetary migration and gravitational scattering events can significantly mix material and/or move planets large distances from their birth orbit. The disk sets the final orbital architecture of the system that sets orbital oscillation amplitudes and frequencies that then affect tidal heating and rotational braking, ultimately altering both internal and atmospheric properties.

The central star provides the largest source of energy in a planetary system and is the dominant influence on the atmosphere. The star's composition can be extrapolated to the disk's, thereby constraining planetary composition (Bond et al. 2010). A star's high energy radiation and stellar wind can alter an atmosphere's composition, possibly even removing it. The host star's gravity can induce a torque on the planet that changes its rotation rate, and an evolving gravitational gradient across its diameter can result in tidal flexing that heats the interior. The star's galactic orbit and composition can constrain its birth location (Loebman et al., 2016).

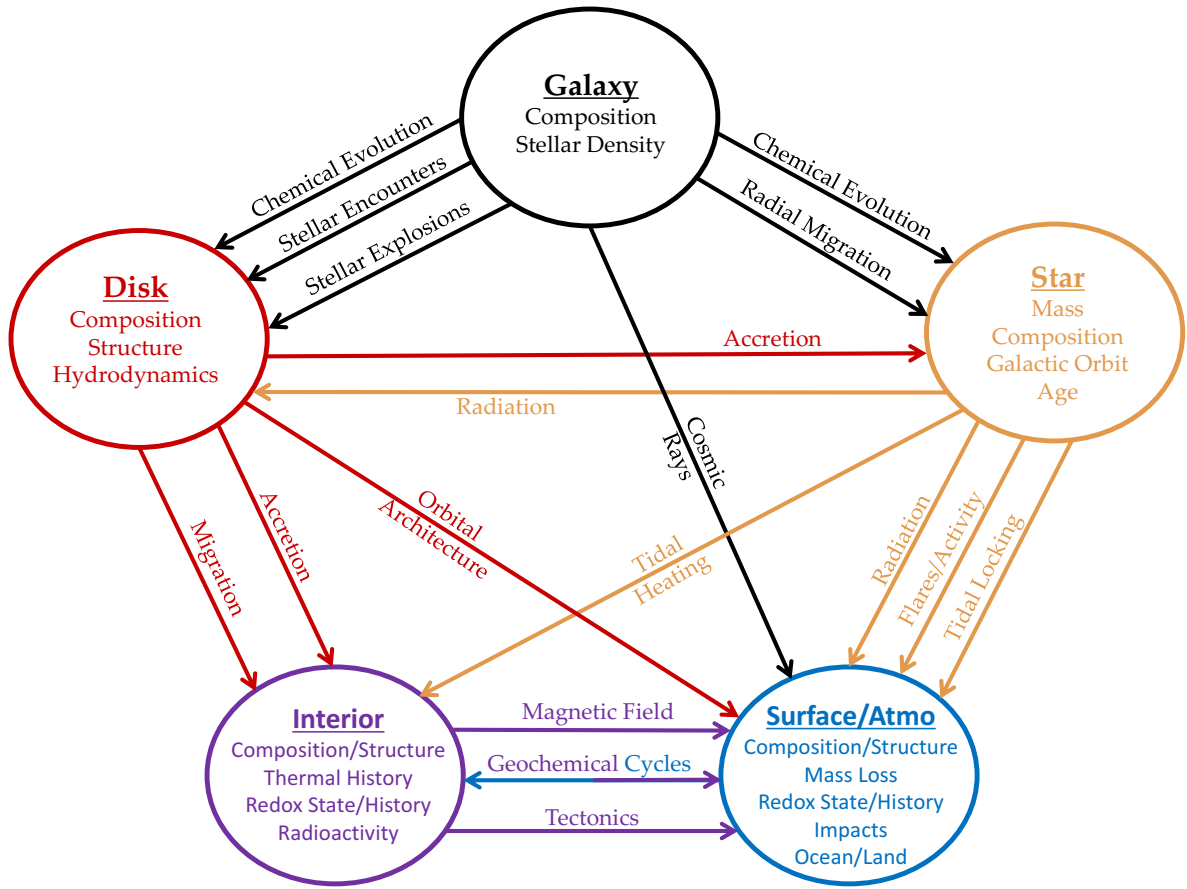

**Figure 1.** Schematic of connections between physical spaces (circles) and processes (arrows) that affect the compositional, thermal, and chemical evolution of terrestrial planets. Exogeoscientists must connect all these phenomena if we are to discover life on exoplanets.

The atmosphere is generated by the stellar heating of surface volatiles into the vapor phase, and by outgassing of internal volatiles. Some chemical species may be modified by stellar light (photochemistry) and others take part in chemical reactions with surface materials (geochemistry), often in complex feedback loops. The surface may be partially or completely covered in liquids. Gravitational perturbations from other planets may drive rotational and climatic cycles that change atmospheric and surface constituents.

The interior has a structure, rheology, and tectonic expression that is determined by its composition and thermal evolution. With sufficient convection in the core, a magnetic field is generated that can extend several planetary radii. Outgassing changes atmospheric composition, and eruptions provide new solid surface material for chemical alteration (weathering). Radiogenic heating, determined by primordial abundances and influenced by galactic birth environment, can provide long-lived energy sources after energy from formation has dissipated. Close-in planets may be heated to the point that solid rock is forbidden and the volatiles in the interior and atmosphere are in equilibrium.

## IV. Future Prospects

Twenty years from now, spectra of hundreds of terrestrial exoplanets will be available, and the scientific community must be prepared to maximally leverage these data to constrain the planets' geophysical and geochemical processes. Ultimately, our ability to infer the formation and accumulation of biosignatures at the surfaces of these worlds – and especially in their atmospheres – will depend on our ability to compare observations with quantitative models that place constraints on the *non-biological* rates of production and destruction of these signatures. This approach can resolve the inevitable debates about "false positives", the non-biological production of putative biosignatures, as well as to avoid sinking precious observing time into planets that are likely to be "false negatives" due to high rates of non-biological consumption. Such models will inevitably be generalizations derived from our understanding of Earth's non-biological geophysics and geochemistry – an understanding that is still incomplete. Any claim for the discovery of life on an exoplanet must be predicated on a mature and robust exogeoscience, anchored to an advanced Earth system science, that can address a huge range of abiotic processes and scenarios.

Therefore, the new exogeoscience community must consist of isotope geochemists, atmospheric spectroscopists, geophysicists, planetary scientists, and galactic astronomers with expertise in observational, experimental, and theoretical methods. Success hinges on institutional support of next generation facilities and strong international collaborations. The interdisciplinary nature of the problem demands opportunities for researchers to share and synthesize ideas. Research consortia, such as NASA's NExSS, should be established to connect these communities.

Exogeoscientists must engage galactic astronomers to elucidate the roles of stellar migration, supernova, and kilonova. Reliably tracing a star's composition to its birth environment (e.g. Loebman et al. 2016) can constrain composition and the likelihood of stellar encounters (Kaib et al., 2013). Models of supernova should be improved to calculate probabilities for the injection of short- and long-lived radionuclides into protoplanetary disks (e.g. Lichtenberg et al., 2016; Fatuzzo & Adams, 2016). The new era of gravitational wave astronomy must be exploited to determine how heavy elements are produced and distributed in the galaxy.

Theoretical models must be capable of simulating plausible formation scenarios to generate initial conditions for planetary system evolution codes that can tractably predict billions of years of evolution. Formation models that self-consistently simulate the collapse of a molecular cloud into a star+disk and the subsequent evolution of that disk are still beyond current technology, but they will be essential for predicting the initial compositions and temperatures of terrestrial exoplanets. Tectonic, geochemical, and magnetic dynamo models must predict atmospheric composition and structure for arbitrary compositions and evolutionary paths. Simulating all the relevant processes will likely take hundreds of scientists and billions of hours of CPU time. Connections with new "big data" methods to handle large data volumes and high dimensional problems will likely be vital for success.

Extrapolating telescopic observations of exoplanetary atmospheres to surface conditions, crustal composition and, ultimately, habitability will require a quantitative understanding of volatile element cycling between the surfaces and interiors of these worlds. But almost all relevant experimental investigations at pressures and temperatures appropriate for planetary interiors have focused on the compositions, temperatures, and pressures of Solar System terrestrial planets (Earth abundances; <6000 K; <360 GPa). Exoplanets likely span a large compositional range and reach much higher pressures and temperatures (~4000 GPa and up to 10,000 K; Duffy et al. 2015). Based on current methodologies, a single researcher will require at least 2 years to analyze one mineral or magma composition at different pressure and temperature conditions, suggesting a thorough exploration will require dozens of people dedicated to the task for the next two decades. In order to accomplish these experiments, it is imperative they have access to a large range of experimental apparati, ranging from low-pressure piston cylinder and multi-anvil devices to high-pressure diamond anvil cells and high-powered laser-driven shock facilities. While the former three instruments are common in laboratories today, the only way to reach pressures higher than 1000 GPa is through shock compression at high-powered laser facilities, such as the magnetic pulsed power Z facility and high-powered Omega laser facilities.

The geochemical evolution of a terrestrial exoplanet over billions of years is essential information for data interpretation, but generating this timeline will be very difficult. A generalized framework for the geochemical evolution of terrestrial planets must be conceived and validated within the Solar System before being applied to exoplanets. Geochemists should be encouraged to record and even publish their "failed" ideas about Earth: Although a mechanism might not occur on Earth today because, say, its carbon abundance is too low, an exoplanet may possess that abundance and the hypothesized process may facilitate the correct interpretation of an exoplanet observation. Importantly, very little work has been done to explore the mineralogy and rock types that can be formed from melts with non-Earth compositions. Experiments that link elemental compositions, volatile solubility, and the mineralogy of mantles and crusts will be critical for establishing the supply of bioessential elements to a planet's surface.

Geophysicists must pursue a generalized understanding of plate tectonics to develop a predictive, dynamic model. This new model must then be connected with geochemical models to understand feedback loops beyond the carbonate-silicate cycle. Weathering of crustal rocks is the key source of many bioessential elements on Earth and is likely to be an important process for exoplanets. The role of the magnetic field must also be resolved so that its presence, or lack thereof, on an exoplanet can further discriminate between sterile and biotic planets. Exo-magnetic fields will offer a unique view into the structure and dynamics of the deep interior, if and when techniques to detect them remotely become available (Driscoll & Olson, 2011).

Finally, we note that systems like TRAPPIST-1 (Gillon et al. 2017; Luger et al. 2017) could be valuable laboratories for exogeoscientists. If terrestrial (see Unterborn et al., 2017b), the inner planets are likely tidally heated, with eccentricities maintained by planet-planet perturbations, and hence are volcanically active and the atmsopheric composition traces internal composition and geophysics. Spectroscopic characterization of their atmospheres accompanied with models of highly volcanic worlds, may reveal the composition of the (less volcanically active) planets in the habitable zone. Systems with "super-Ios" and habitable zone planets may offer the best opportunity to understand the exogeoscience of potentially habitable exoplanets. Observations of such a system, accompanied with a robust exogeoscience analysis, may offer the fastest route for the detection of life on an exoplanet.